\documentstyle[prd,aps]{revtex}
\input epsf.tex
\def\DESepsf(#1 width #2){\epsfxsize=#2 \epsfbox{#1}}
\begin{document}
\draft
\preprint{CTP-TAMU-43-98}
\title{Trilepton Signal of Grand Unified Models at the Tevatron }
\author{E. Accomando, R. Arnowitt and B. Dutta}
\address{Center for Theoretical Physics, Department of Physics, Texas A\&M
University,  College Station, TX 77843-4242}
\date{November, 1998}
\maketitle
\begin{abstract}

At the Tevatron, the most promising channel to detect
supersymmetry is three leptons plus missing energy, where the leptons are
$e$'s and/or $\mu$'s. This final state appears from the production of chargino
and second lighetst neutralino. However in grand unified models with 
universal scalar masses at the grand unified scale, this final state mostly
consists of $\tau$'s which are hard to detect. We show that for some regions
of  non universality in the scalar masses at the GUT scale based on  
unifying groups like SU(5) or SO(10), the final state 
mostly consist of 3$l$+${\rlap/E}_T$ and 
$\tau ll$+${\rlap/E}_T$. The first mode has very high detection efficiency and
the second one is expected to have high detection efficency as well. We also
show that these models can have enough events in these modes to be detected in
RUN II.
\end{abstract}   

\pacs{ }

 \vskip2pc]
Due to its many attractive features, supersymmetry (SUSY) has become
the main focus  of experimental search.  Among the existing colliders, the
upcoming RUN II ($\sqrt s=2$ TeV and 2 $fb^{-1}$luminosity ) of the Tevatron is going to have the highest reach in terms of
scanning the supersymmetric parameter space. Consequently, final states of
different SUSY particles  are being examined on the basis of their
detectibility.  

The production of  the lightest chargino and the second lightest neutralino 
($\chi_1^{\pm},\chi^0_2$)is found to be very promising for the discovery of
SUSY. In most of the supersymmetric models,  for large regions of the parameter
space, the  masses of
$\chi_1^{\pm}\,{\rm and }\,\chi^0_2$ are  within the reach of RUN II. The final
states of this mode can involve three leptons + ${\rlap/E}_T$. If these
trileptons are $e$'s and $\mu$'s, then the detection of the final states becomes
easier since 
$e$, $\mu$ have large detection efficiencies ($\sim 85\%)$. If, however, the
final states involve $\tau$'s, the detection becomes harder, since the $\tau$
detection efficiency has not yet been  specified. The $\tau$ can be detected
hadronically (`thin jet') or leptonically. If we depend on the leptonic modes
(since we are working in a machine with lots of jets), then the effective
leptonic cross  section of the final state with multi $\tau$'s becomes very small
(due to  the small leptonic branching ratio of the $\tau$).

The trilepton channel at the Tevatron has been analysed by many groups in various 
different supersymmetric models \cite{w1z2,gmsb}. These theoretical scenarios
range from the supergravity motivated models (SUGRA) to Gauge Mediated SUSY
breaking Models (GMSB). In absence of experimental evidence as well as a full 
understanding of the dynamics of SUSY breaking, no unique model  has as yet
emerged. Models where the different SUSY masses are related, are found to be more
compelling due to their predictivity. Among these, the SUGRA models with the
constraint that  all the scalar masses are same (and gaugino masses also the
same) at the grand unified theory scale $M_G$ (universal boundary conditions) is the most  popular
one.  The major finding of the trilepton search analysis  in this model is
that the 
$\tau$ dominated final states are more abundant for large as well as for 
smaller $\tan\beta$ \cite{bk1}. GMSB models give rise to hard photons (easy to
detect) or high $p_T$ $\tau$  in the final states of the ($\chi_1^\pm,\chi_2^0$)
production.

In this letter, we will examine the  chargino-neutralino 
($\chi_1^{\pm},\chi^0_2$) pair production in grand  unified models within
the framework of  SUGRA with radiative breaking of the electroweak symmetry. We
will not impose  universality of the scalar masses at $M_G$. Instead, we will
impose the  constraints on the masses based on the representations of the 
groups to which these fields belong and will use unifying groups like SU(5) and
SO(10). We find that these non universal boundary conditions may lead to specific and
detectable characteristics  in the trilepton final state. In parts of the
parameter space the final state infact is
dominated by 3
$l$ +${\rlap/E}_T$($l$ is $e$,
$\mu$) or $\tau ll$+${\rlap/E}_T$ instead of all $\tau$. Due to the high
detection efficiency 3$l$ mode is the best candidate and due to high luminosity, one can have a lower threshold to trigger them via a dilepton
trigger $ll$+${\rlap/E}_T$.  In the same fashion the
$\tau ll$ mode is also expected to have high detection efficiency\cite{tk}. 
 Analysing the trilepton signal in various 
channels in RUN II, this general boundary condition (non-universal scenario) may
be distinguished from the  universal one in wide regions of the parameter space.

In supergravity motivated unified models, the non-universality at the
boundary can appear naturally.
 A general non-flat  Kahler metric (where the SUSY breaking field is coupled to
the observable fields with different couplings)  can induce non-universalities in
the scalar masses \cite{sugra2}.  Since the Higgs sector is weakly constrained by the requirement
of FCNC suppression and the third generation is only weakly
coupled to the FCNC processes, one  may assume that the third generation squark, slepton and
Higgs masses  are non-universal at the GUT scale, while the first and 
second generation scalar masses and the gaugino masses are assumed to be
universal. (Non universalities in the gaugino sector can be also induced, but 
are small in most models, and so we assume these masses to be universal). 
Scalar mass non universalities also can be  generated from the running of the
RGE's from the Planck scale or string scale  to the GUT scale. In this case, 
 because of the quark-lepton
unification, not only the third generation squark masses, 
but the third generation slepton masses will be different from the
other generation masses \cite{hb3}.  Finally, we mention that  non universalities
can be generated from the  so-called D-terms arising from the rank reduction of
the groups which embed the SM as a subgroup, when the GUT group has  ranks
higher than the SM .  

Let us examine the parametrization of the  non-universalities. The Higgs soft
breaking masses are given by $ m_{H_1}^2=m_0^2(1+\delta_1)$;
$m_{H_2}^2=m_0^2(1+\delta_2)$.
The third generation fermion soft breaking masses are as follows:
$m_{q_L}^2=m_0^2(1+\delta_3)$; $m_{u_R}^2=m_0^2(1+\delta_4)$
$m_{e_R}^2=m_0^2(1+\delta_5)$;
$m_{d_R}^2=m_0^2(1+\delta_6)$;$m_{l_L}^2=m_0^2(1+\delta_{7})$.
 The $\delta_i$ exhibit the amount of non universality. If we use
a unifying group, where the fields belong to some represantaion of that group the
$\delta_i$ develop relations among themselves. For example, in the case of
the GUT group SU(5) the matter fields are embedded in $\bar 5$ and $10$
representations. The $\delta_i$ in the previous expressions have the following
relations:
$
\delta_3=\delta_4=\delta_5=\delta_{10};~~\delta_6=\delta_7=\delta_{\bar5}.
$ 

Any GUT group which has an SU(5) with the matter
fields in the
$10$ and $\bar 5$ representations will have the above pattern of
non universalities. In the case of  $SO(10)$, if we demand a direct breaking  into
the SM and keep the $5+\bar 5$ Higgs in the same 10 of $SO(10)$, we get an
additional constraint, $\delta_5=\delta_2-\delta_1$. In this note we will use
$\delta_{1}$,$\delta_{2}$,
$\delta_{\bar5}$ and
$\delta_{10}$ to represent the nonuniversalities.  One need not be restricted in
this choice.

In order to determine the physical masses we need to know $\mu$ which is
determined from the electroweak symmetry breaking condition. For
tan$\beta$ small enough, one obtains the following  analytical expression
\cite{mu5}: $
\mu^2 = \mu^2_{\rm univ} + {m^2_0\over{t^2-1}}(\delta_1-\delta_2 t^2-{D_0-1\over
2} (\delta_2+2 \delta_{10})t^2)+{3\over 5}{t^2+1\over{t^2-1}}S_0p +
{\rm one loop},
$ where $t=\tan\beta$, $D_0\simeq 1-(m_t/200\sin\beta)^2$,
$S_0=Tr(Ym^2)$,  p=0.0446 and $\mu^2_{\rm univ}$ is the remaining
 universal part.

Let us first  discuss the trilepton signals in the case of universal boundary
conditions. It has been shown that\cite{bk1} among all the trilepton plus
${\rlap/E}_T$ modes, only the 3$\tau$+${\rlap/E}_T$ production cross section is
large in most of the parameter space when the sleptons are produced on shell. 
Among the other modes, only $\tau ll$ becomes comparable
but only for the very small region of $\tan\beta$ $\le 5$.  The branching ratio
(BR) in this mode
 decreases rapidly and at $\tan\beta$=10, for e.g. 
 $m_0=100$ GeV and $m_{1/2}=200$ GeV, the $\tau ll$ mode
  becomes about $1\over {10}$ of the BR of 3$\tau$ mode (The  BR of 3$\tau$
remains almost the same). The $\tau ll$ cross section becomes 21.2 fb. The
situation worsens by decreasing $m_0$. For example, at $m_0=30$ GeV (keeping the
other parameters the same) the cross section for the $\tau l l$ mode  becomes 11 fb
(and in this region the BR of $\tau\tau l$ becomes large). 
Even when the sleptons
become offshell, the branching ratio to  the leptonic modes involving multi
$\tau$'s is large.  The $\tau $ domination in the signal (onshell or offshell
case) is usually expected when $\tan\beta$ is large.   But the domination seems
to persist even in the region of low $\tan\beta$.

The reason for this misfortune depends primarily on two factors. The lighter 
stau ($\tilde\tau_1$) mass is lighter than the selectron mass. In the onshell
case, the decay width depends on  $(\Delta m^2)^2$(where $\Delta m^2$ is the
mass$^2$ difference between the gaugino and the slepton), which  is larger for
the modes involving
$\tau$.  The other factor is that the $\chi^0_2$
 is primarily a wino which has coupling to the left sleptons only which
  are  heavier than both the right handed selectrons and 
the $\chi^0_2$. On the otherhand,  the lighter
$\tilde\tau_1$ is a mixture of $\tilde\tau_L$ and $\tilde\tau_R$ due to the 
large left right mixing $m_{\tau}\mu
\tan\beta$. (We will assume $A=0$ at the GUT scale  throughout the analysis, but
a non-zero value will not change the conclusion). Consequently, $\chi^0_2$
primarily will decay into the lighter
$\tilde\tau_1$ and a $\tau$.  Among these two
 factors, the first one has a larger impact. The $\chi^{\pm}_1$ is a mixture of
charged Higgsino and  wino giving rise to a dominating  $\tilde\tau_1\nu$ final
state when the $\tilde\tau_1$ is onshell. 

In the case of non-universal boundary conditions,  new effects 
can reduce the right-handed selectron mass and raise the  $\tilde\tau_1$ mass
(and can make it even larger than the selectron mass) and  the BR into the 3$l$ 
mode is no longer suppressed.  The magnitude of $\mu^2$  can also be decreased
which 
 increases the $\tilde\tau_1$ mass and thereby decreases the BR into $\tau$
dominated final
 states. Finally, the nature of 
$\chi^0_2$ can change with the change in the size  of $\mu$ (the wino component
can decrease and the bino component can  increase).

Another important point to note is that, when we use the non universal boundary
conditions we have to add the term 
$S\equiv \alpha_1{{3 Y_i}\over {10 \pi}}\sum_i(Y_i m_i^2)$ to the RGEs of the
fields. The contribution from this term is zero in the case of universal
boundary condition but is non zero in the  non-universal
case. 

We are now ready to discuss the results.   In Fig.1 we show the  masses
 as a function of the nonuniversalities. In Fig.1a we plot the
$\chi^\pm_1$, $\tilde\tau_1$,
$\tilde\nu_L$, $\tilde e_R$ and $\mu$ as functions of $\delta_{10}$. The other
$\delta_i$ are  zero. We see that the  $\tilde\tau_1$ mass increases as
$\delta_{10}$ increases. Since
$\delta_{10}$ does not contribute to the S term, the $\tilde e_R$ or 
$\tilde\nu_L$
masses do not get any effects. The $\chi^0_2$ mass is very close to
the $\chi^{\pm}$ mass. 

In Fig. 1b we plot the the same masses as functions of $\delta_2$. Since
$\delta_2$ contributes to the S term, we see that the $\tilde e_R$ mass
decreases for the positive values of $\delta_2$. The $\tilde\tau_1$ mass also
decreases up to moderate values of $tan\beta$, but the decrease is  slightly
lower than the $\tilde e_R$ mass which helps to raise the raise the $3l$ or
$\tau ll$ BRs.  This happens because, as
$\delta_2$ increases, $\mu$ gets reduced which reduces the off
diagonal element in the stau mass matrix. The reduction in the size of $\mu$
also lowers the
$\chi^{\pm}_1\,{\rm and}\,\chi^{0}_2$ masses. In
the large
$tan\beta$ case ($tan\beta\stackrel{>}{\sim} 25$,
 depending on the size of $\delta_2$) the
$\tilde\tau_1$ mass is increased which helps to increase the 3$l$ or $\tau ll$
BRs.

  In the Figs.2 and 3, we plot the production cross sections of the leptonic
modes  as function of $m_0$. We have chosen a pattern of non-universality which 
obeys the
$SU(5)$ group structure.
 We have used
$\delta_{1}=-0.5$,
$\delta_{2}=0.5$,
 $\delta_{\bar 5}=1$ and $\delta _{{10}}=0.5$. These values of the $\delta_i$ 
also allow the simplest SO(10)  breaking patterns (SO(10)$\rightarrow$SM) at $M_G$  i.e. the condition
$\delta_{\bar5}=\delta_2-\delta_1$ is satisfied.
 
In Fig.2a, we use $\tan\beta=10$,
$m_{1/2}=200$ GeV and $\mu>0$.  The sign convention we adopt is the same  as in
ref\cite{hk6}.  In Fig.2b we plot the same parameter  space for the universal
boundary condition. We observe the following:

\noindent a) For $30\stackrel{<}{\sim} m_0\stackrel{<}{\sim} 70$ GeV, the 3$\tau$ and $\tau\tau l$ mode dominate 
initially in the non-universal case. But the $\tau ll$ mode is not far behind and as 
$m_0$ increases the BR in this mode increases. The reason for $\tau$  domination
 in  this region (the $\tilde e_R$ and $\tilde\tau_1$ masses are almost same) is
due to the Wino nature of $\chi^0_2$. In this region the  chargino can decay into
$l\tilde\nu_l$. As the sneutrino mass goes  offshell towards the end of the
region, the branching ratios of the 3$l$ and
the $\tau\tau l$ modes reduce. In the universal case (Fig 2b), the 3$\tau$ mode dominates
with the $\tau ll$  and the $\tau\tau l$ modes coming next.

\noindent b) For $70\stackrel{<}{\sim} m_0\stackrel{<}{\sim} 100$ GeV, the $\tau ll$ and 3$\tau$ mode dominate in
Fig.2a,  and the 3$l$ mode starts becoming significant. In this region the 
selectron mass becomes lower than the $\tilde\tau_1$ mass and hence the BR of 
$\chi^0_2$ into $e$'s and $\mu$'s increases. Towards the end of the region, the  sleptons
(first $\tilde\tau_1$) becomes offshell. In the universal case the  3$\tau$ mode
is dominant in this region.
 
\noindent c) For $100\stackrel{<}{\sim} m_0\stackrel{<}{\sim} 200$ GeV, the sleptons are mostly offshell. In Fig.2a
the 3$l$ is the dominant decay mode and next to that is the $\tau ll$ mode. The
production cross section decreases as we increase $m_0$.  At the end of this region the 3$\tau$ and 
$\tau\tau l$ modes increase again due to the offshell Higgs contribution. In the
universal case the 3$\tau$ mode dominates initially. For $m_0\ge 130$, the 3$l$ 
mode becomes equal to the 3$\tau$ mode, but the cross section is very reduced by
that time. (The $\tau\tau l$ mode dominates here).

Using a dilepton trigger, and if we use 5  events in RUN II as a bench mark for a SUSY
signal (corresponding to the cross section of 25 fb with 10$\%$ 
acceptance rate \cite{cdf}
), the inclusive $ll$+${\rlap/E}_T$ production allow us to scan 
$50\stackrel{<}{\sim} m_0\stackrel{<}{\sim} 130$ GeV. In the universal case we do not get 
5 events in this mode for any value of $m_0$. In the case of a $\tau$l trigger, 
(assuming the same acceptance rate) the inclusive $\tau l$+${\rlap/E}_T$ 
production allows us to scan $30\stackrel{<}{\sim} m_0\stackrel{<}{\sim} 100$ GeV. 
Here again, in the universal case we do not find any $m_0$ value which gives
rise to 5 events.
 
If we reduce $m_{1/2}$, the BRs of the $\tau ll$ and the 3$l$ mode increase 
more because of the increase in phase space. 
 In Fig.3 we plot the  cross sections of the leptonic modes as function of $m_0$
for
$m_{1/2}=150$ GeV and $\mu>0$. 

In the case of $\mu<0$, the
$\chi^{\pm}_1\chi^0_2$ production cross section decreases due to increased
chargino and  neutralino masses.  We find that this choice of the 
$\mu$ sign gives rise to
a scenario  analogous to what one could get by increasing the value of
$m_{1/2}$. The contribution of the on shell sneutrino for $m_0\stackrel{<}{\sim}
 80$ GeV ($m_{1/2}=200$ GeV)
enhances the  branching of the chargino into leptons. The 3$\tau$ channel
shrinks. 

We have discussed above the dependence of the leptonic signal on 
$m_{1/2}$ and $m_0$ at a fixed value of $tan\beta$ and for a particular  set of
$\delta_i$. We next examine the correlation between $tan\beta$  and the
non-universal boundary conditions and find that, even for  moderate to large
values of $tan\beta$, the abundance of $\tau$'s in the  final state can be
reduced in favor of the 3$l$ and  $\tau ll$ channels for positive $\delta_{10}$
and $\delta_5$.
  
 Fig.4 shows a parametric plot of the 3$l$ and $\tau ll$ with the cross 
section fixed at 35 fb as functions of  $\delta_{10}$ and $tan\beta$. Here  we
have used $\delta_2=0.5$, $\delta_1=-0.5$, $\delta_5$=1, $m_0=100$ GeV,
$m_{1/2}=200$ GeV and $\mu>0$.  Both $\delta _5$ and $\delta _{10}$ help to
raise $\tilde\tau_1$ mass. Since
$\delta_{10}$ affects the right stau mass, it has a larger impact in increasing 
$\tilde\tau_1$ mass and thereby decreasing the branching ratio of the
$\tau\tau\ l$ and 3$\tau$ modes. The selectron mass is reduced by a small
amount  (a few GeV) through the S term.  Hence an increment of $\delta_{10}$
will raise the BR of the 3$l$ and $\tau ll$ mode. It is evident from the figure
that as
$\tan\beta$  increases, larger values of $\delta_{10}$ are needed in order to
compensate  the decreasing of the $\tilde\tau_1$ mass. However, one sees that
for $\delta_{10}\leq 2$, the 3$l$ mode is large even for
$tan\beta\simeq 25$.  

 An increase in the magnitude of $\delta_2$ raises the 3$l$ and 2$l$ BRs. 
 (The bino
component of $\chi^0_2$ increases.)  But since the production cross section
becomes small  (the coupling becomes smaller because the wino component of
the $\chi^0_2$
   reduces), the net increase in the cross
sections of the same modes is small. Hence a change in $\delta_2$ does not affect
the cross sections of the leptonic modes much.

One can use other decay chains of SO(10), e.g.
$SO(10)\rightarrow SU(3)_c\times SU(2)_L\times SU(2)_R\times U(1)_{B-L}$. In
this case again the $\tilde \tau_1$ mass can be increased by increasing the
$m^2_{\tilde \tau R}$ mass  by adding a non universality.
This term can affect the $\tilde e_R$ through the S term, but the effect is
small. One can also cancel this effect by adding an identical magnitude of non
universality to  
$m^2_{\tilde \tau L}$. The net effct is the $\tilde \tau_1$ will increase further
without changing the $\tilde e_R$ mass.  

To conclude, we have looked into the final states of the chargino-second
lightest neutralino production at the Tevatron for $\sqrt s=2$ TeV.  
In the models
with universal boundary conditions the 3$\tau$ +${\rlap/E}_T$ mode dominates
among all the leptonic final states even for low values of $\tan\beta$. 
We have 
found that in grand unified models with non universal boundary conditions, 
the
3$l$ and $\tau l l$ final states can dominate over the 3$\tau$ or the $\tau\tau
l$ modes for low and intermediate values of
$\tan\beta$ for some regions of non universality. The magnitudes of the 
non universalities can be relatively small in order to give rise to the  
above
type of scanarios.  This result holds for either sign of $\mu$. 
 Since the 3$l$ mode has by far the best detection efficiency, these
nonuniversal boundary conditions may be tested in RUN II.

Work of E. A. and R. A  have been supported by the National Science 
Foundation grant No. 9722090
 We thank  Teruki Kamon for his critical comments and suggestions. We also thank J.P. Done for useful discussions.
 
\newpage
\begin{figure}
\centerline{ \DESepsf(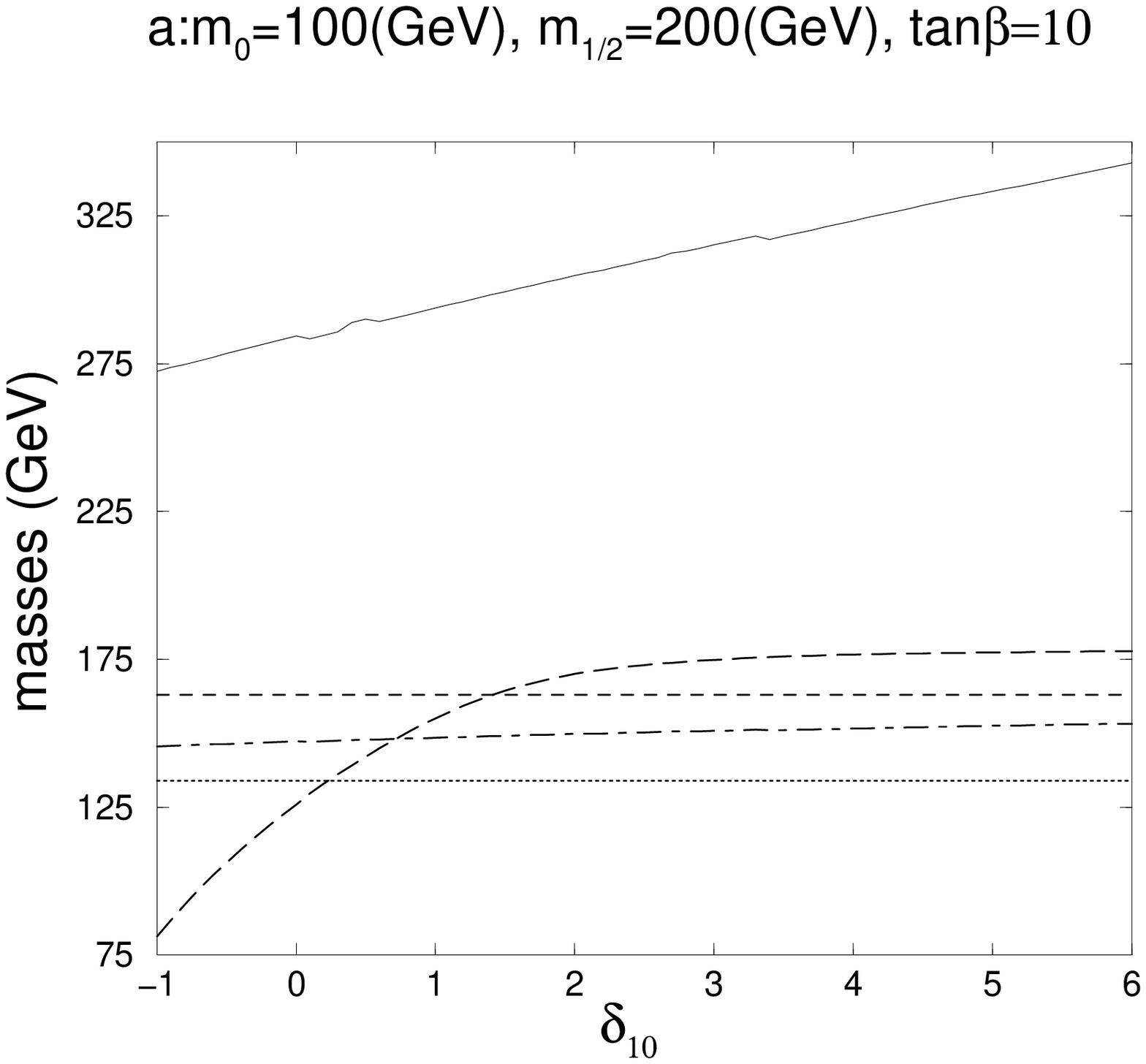 width 7 cm) }
\centerline{ \DESepsf(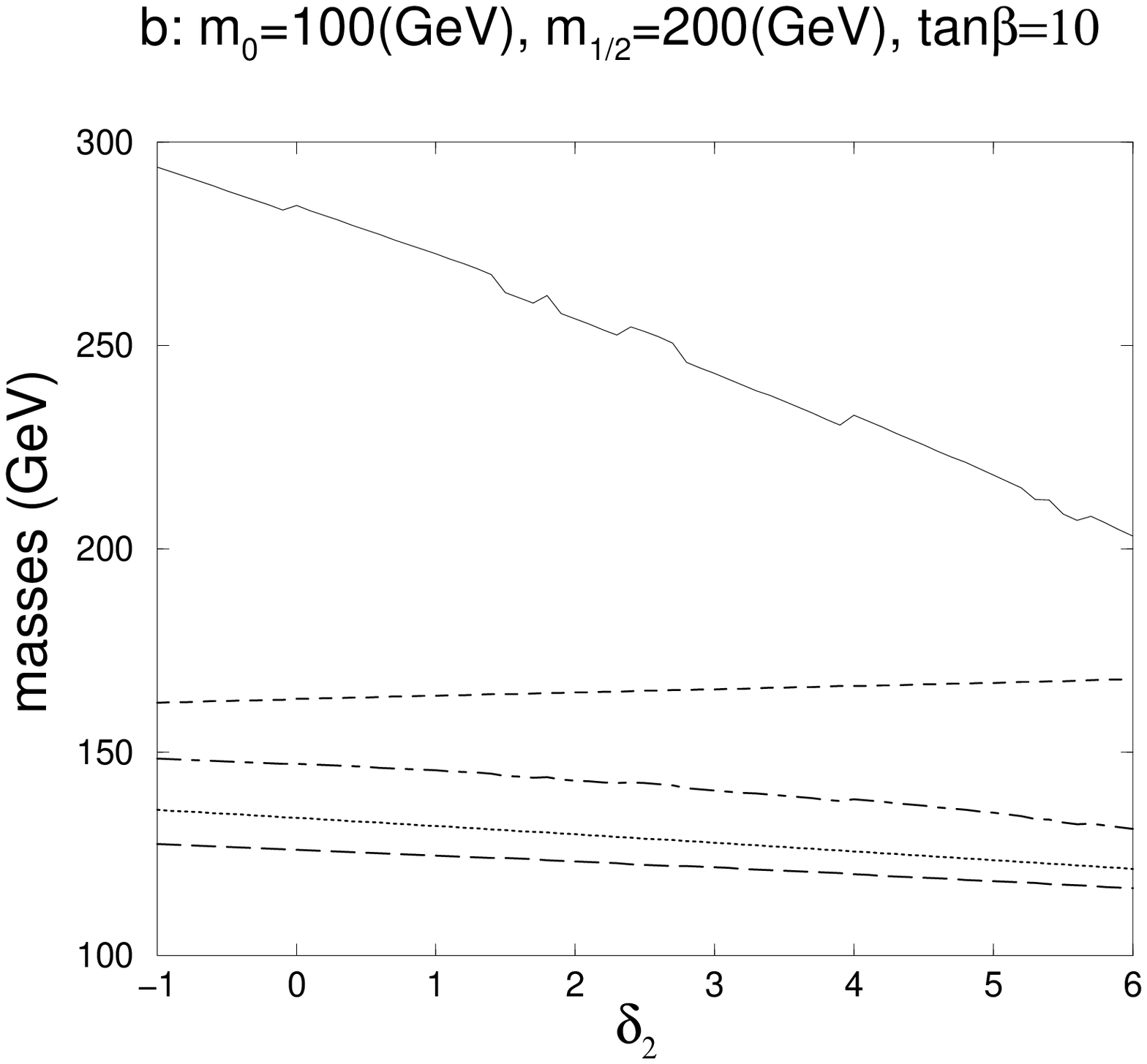 width 7 cm) }
\caption{ $\mu$ (solid line), $\tilde\tau_1$ (long dashed), $\tilde\nu_L$
(short dashed), $\tilde e_R$ (dotted) and $\chi^\pm_1$ (dot dashed) are shown }
\end{figure}
\newpage
\begin{figure}
\centerline{ \DESepsf(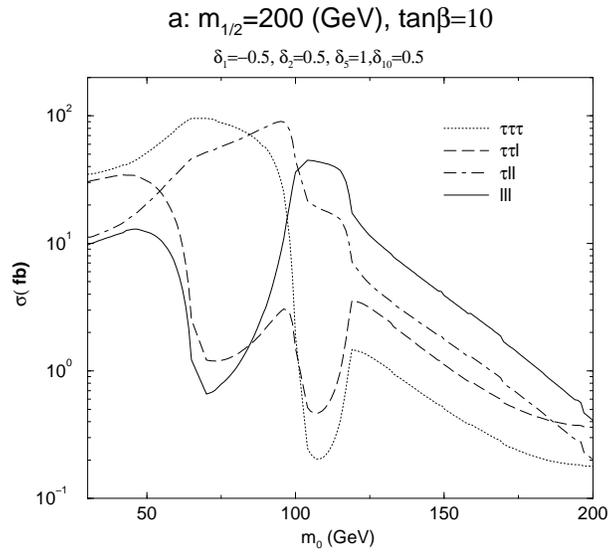 width 8 cm) }
\centerline{ \DESepsf(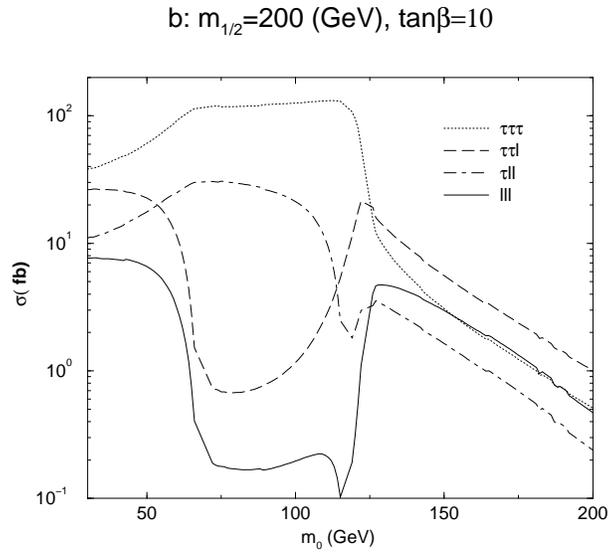 width 8 cm) }
\caption{Cross sections for all the leptonic modes.
a)Non-universal case and b)universal case.}
\end{figure}
\newpage
\begin{figure}
\centerline{ \DESepsf(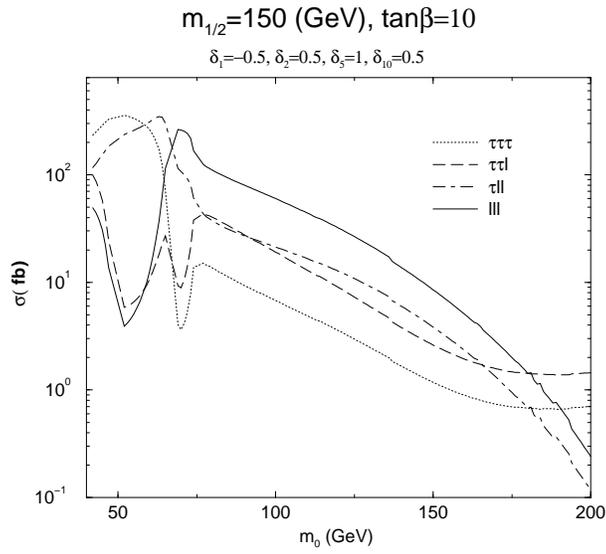 width 8 cm) }
\caption{Cross sections for all leptonic modesfor the
non-universal case}
\end{figure}
\begin{figure}
\centerline{ \DESepsf(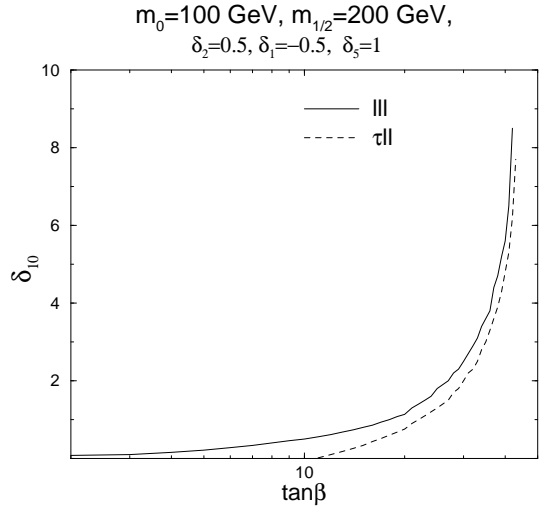 width 7 cm) }
\caption{The value of $\delta_{10}$ required as a fuction of $tan\beta$  so that
the production cross section for the 3$l$ and  the $\tau ll$ modes be $35$fb.}
\end{figure}
\end{document}